\documentclass[twocolumn]{aastex631}
\usepackage{booktabs}
\usepackage{graphicx} 
\usepackage{natbib}
\usepackage{orcidlink}
\usepackage{hyperref}

\begin{document}

\title{Rotational Light-Curve Recovery \& Predictions of the LSST Yield of Hildas}

\author[0009-0008-2687-0422]{Alexander J. Fleming}
\affiliation{Department of Astronomy, University of Washington, Box 351580, Seattle, WA 98195-1580, USA}
\email{aflem110@uw.edu}

\author[0009-0005-5452-0671]{Jacob A. Kurlander} 
\affiliation{DiRAC Institute \& the Department of Astronomy, University of Washington, 3910 15th Ave NE, Seattle, WA 98195, USA}

\author[0009-0007-1972-5975]{Dmitrii E. Vavilov}
\affiliation{DiRAC Institute \& the Department of Astronomy, University of Washington, 3910 15th Ave NE, Seattle, WA 98195, USA}

\author[0000-0002-6034-5452]{David Vokrouhlický}
\affiliation{Astronomical Institute, Charles University, V Holešovičkách 2, CZ 18000, Prague 8, Czech Republic}

\author[0000-0002-4547-4301]{David Nesvorný}
\affiliation{Department of Space Studies, Southwest Research Institute, 1301 Walnut Street, Suite 400, Boulder, CO 80302, USA}

\author[0000-0003-0743-9422]{Pedro H. Bernardinelli}
\affiliation{DiRAC Institute \& the Department of Astronomy, University of Washington, 3910 15th Ave NE, Seattle, WA 98195, USA}

\author[0000-0003-1996-9252]{Mario Juri{\'c}}
\affiliation{DiRAC Institute \& the Department of Astronomy, University of Washington, 3910 15th Ave NE, Seattle, WA 98195, USA}

\begin{abstract}

    The Hilda population occupies the stable 3:2 mean-motion resonance of Jupiter and provides a window into Solar System evolution, including collisional processes. The NSF-DOE Vera C. Rubin Observatory will conduct the ten-year Legacy Survey of Space and Time (LSST). We present a simulation of Rubin's discovery of Hildas with the \texttt{Sorcha} \citep{Sorcha_steph, Sorcha_matt} survey simulator and the recovery of their light curves. We constructed a synthetic Hilda population model which includes distributions of orbital properties, sizes, collisional families and colors. We included two color classes corresponding to the Jupiter Trojan populations \citep{wong_17_trojan_colors}. We applied three distinct populations of sinusoidal light-curves to this same orbit–size–color model: (1) a Gaussian kernel density estimate (KDE) fit to rotational periods and amplitudes from the Lightcurve Database \citep[LCDB;][]{LCDB} (2) a super-fast rotator (SFR) population (0–3 hours) and (3) a super-slow rotator (SSR) population (100–1400 hours). Over the ten-year simulated survey, we predict LSST will discover {$\sim$}33,400 Hildas, a fivefold increase over the known population. Using a multiband Lomb-Scargle Periodogram via \texttt{Astropy} \citep{astropy:2022} we confidently recover {$\sim$}46.5\% of Hildas in our LCDB-based population, higher than typical in observational searches. This suggests our light-curve population model may differ from the intrinsic population. We find strong biases in light-curve amplitude, with recovery efficiency dropping sharply below 0.1 magnitudes, while biases from rotational period are comparatively weak aside from cadence-related features such as LSST’s $\sim$36 minute revisit cadence. Our recovery efficiency is likely overestimated due to our assumption of constant sinusoidal light-curves, which correspond to optimal pole orientations. These results are the first test of light-curve recovery from simulated LSST observations. 

\end{abstract}

\section{Introduction}

\begin{figure*}
    \begin{center}
        \hspace{10mm}
        \includegraphics[height=0.35\textheight,width=1\linewidth]{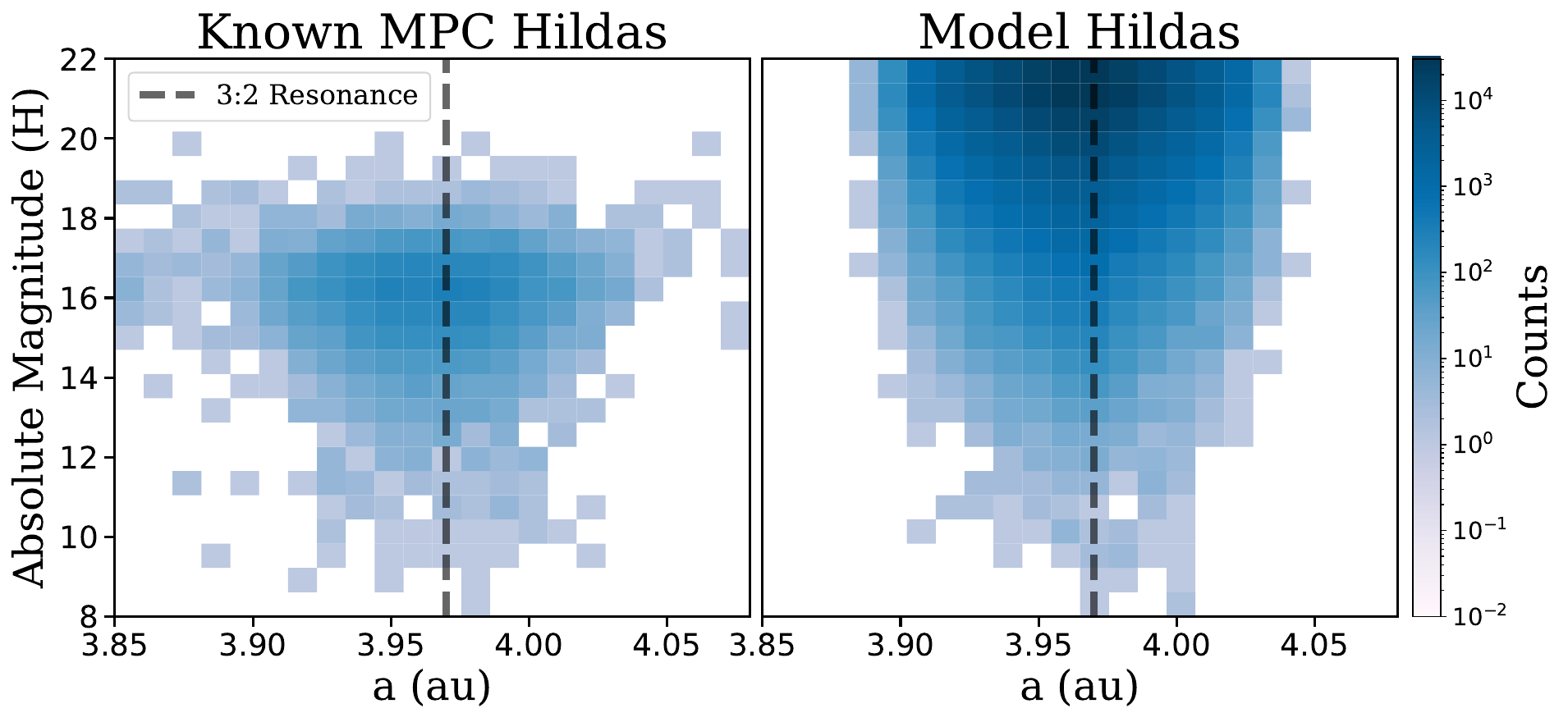}
    
        \caption{Comparison of the known Hilda population from the Minor Planet Center (left) and our synthetic Hilda population (right) in semi-major axis ($a$) and absolute magnitude ($H$). The vertical line at $a = 3.971$ AU marks the location of the 3:2 mean-motion resonance with Jupiter, which defines the Hilda population. Our input sample reproduces the observed distribution in both $a$ and $H$, while extending to fainter magnitudes beyond the MPC limit.}

        \label{fig:mpchist}
    \end{center}
\end{figure*}

\begin{figure*}
    \begin{center}
        \centerline{\includegraphics[height=0.35\textheight,width=.75\linewidth]{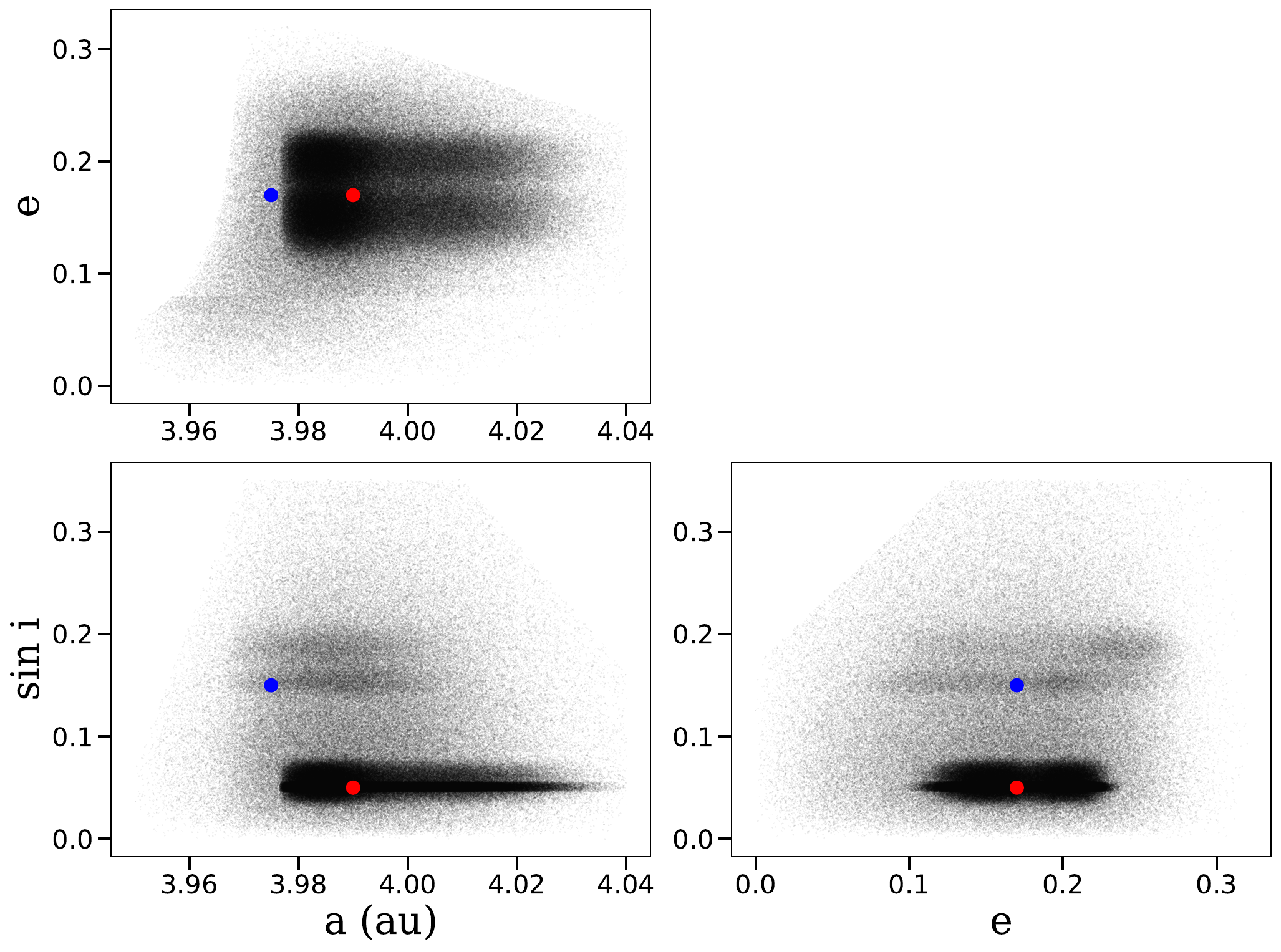}}
    
        \caption{Distribution of our 485,807 simulated Hildas projected onto 2D planes of their \textit{proper} orbital parameters. (i) (a, e) top left; (ii) (a, $\sin (i)$) bottom left; and (iii) (e, $\sin (i)$) bottom right. Each dark point represents an object from our simulated Hilda population. Approximate locations of the two largest collisional families associated with (153) Hilda (blue marker) and (1911) Schubart (red marker) are indicated for reference \citep{david&david}.}

        \label{fig:2d_orbital_elements}
    \end{center}
\end{figure*}

The Hildas are a dynamically stable population of small bodies in the 3:2 mean-motion resonance of Jupiter, with semi-major axes near $\sim 4$~au. Their resonance prevents close encounters with Jupiter and preserves a characteristic triangular configuration in the frame co-rotating with Jupiter \citep{CC_orbital_light_curves}. Because of their long-term stability, Hildas preserve clues about planetary migration and Solar System evolution \citep{david&david} such as the timescales over which the outer planets reached their current orbits \citep{fra2004,morbidelli_asteroid_evo} and the epoch when this migration occurred, whether during giant planet instability or earlier in the primordial Solar System \citep{fra2004,lev2009,rn2015,vok2016,nes2018,pir2019}. Their absolute magnitude (and thus inferred size) distributions have been characterized by \citet{david&david}~(Figure \ref{fig:mpchist}), who found that the background population and collisional families exhibit distinct magnitude-distribution slopes. Their optical colors closely resemble those of the Jupiter Trojans, exhibiting a similar bimodality \citep{wong_17_trojan_colors}. We drew our colors from D-type asteroids and hint at a possible common origin with the Jupiter Trojans and outer Solar System bodies, making the Hildas the nearest large reservoir of trans-Neptunian material in the Solar System. Roughly 60\% of currently known Hildas are associated with collisional families \citep{bv2008,david&david} and this fraction may vary with object size due to differences in collisional size distributions. Figure~\ref{fig:2d_orbital_elements} shows our simulated Hilda population in the space of proper orbital elements, following the approach of Figure 6 in \citet{david&david}, and highlights the clusters corresponding to the major collisional families. These collisional families (from largest to smallest) are Hilda family, Schubart family and Potomac family. Studying these families can provide insight into collisional physics in the outer main belt, the evolution of fragment orbits due to non-gravitational effects like the YORP effect, and the long-term stability of asteroid populations. 

As of October 15, 2025, the Minor Planet Center (MPC) lists 6,622 known Hildas \citep{MPC_db_search}, mostly discovered incidentally by major asteroid surveys rather than dedicated Hilda searches. The Vera C. Rubin Observatory’s Legacy Survey of Space and Time (LSST) represents the next major leap in Solar System science. Its ten-year duration, six-filter optical/NIR coverage, image depth and broad sky coverage will enable unprecedented characterization of the Hilda population, including their rotational properties. One of LSST's primary scientific goals is to obtain an ``inventory of the Solar System" \citep{LSST2009} by discovering and characterizing as many small bodies as possible. Using a comprehensive simulation of LSST's Solar System capabilities, \citet{jake_lsst} found that LSST will discover several times more objects than are currently known in each orbital class and will meet the Rubin Observatory Metrics Analysis Framework \citep{MAF} metric for high-quality light curves on 3–8\% of the non-NEO bodies it observes. Given their comparable dynamical states, one might expect similar behavior between the Hilda and Trojan populations, but this expectation remains untested as no equivalent LSST yield analysis has been conducted for the Hildas. 

We use \texttt{Sorcha} \citep{Sorcha_steph, Sorcha_matt}, a Solar System survey simulator that models LSST's detection and discovery of asteroids and creates a simulated source catalog. We then perform our own rotational analyses on these simulated detections to assess LSST’s ability to measure rotational properties. Simulating the survey enables us to examine how cadence, depth and other observational factors influence which objects are detected and how well their light curves are sampled. These detection and effects then determine which rotation periods and amplitudes can be reliably recovered, making it essential to quantify their impact when interpreting the observed light-curve distributions and inferring the intrinsic properties of the Hilda population. 

Rotational light curves are the primary means of determining asteroid spin rates, shapes and surface features, but they are challenging to obtain. Of the 6,622 Hildas currently listed by the Minor Planet Center, only 196 have measured light-curve amplitudes and rotational periods in the Light-Curve Database \citep[LCDB;][as of June 26, 2025]{LCDB} and only about one third of these objects have the full spin state reconstructed \citep[e.g.,][]{dh2023}. This small fraction may reflect observational limitations or intrinsically low variability in the Hilda population if most are axially symmetric. The currently available empirical Hilda population light-curve data is strongly shaped by observational biases and does not represent the intrinsic distributions of rotational period or amplitude. Larger amplitudes are typically easier to find, while successfully finding a rotational period of an object with roughly zero amplitude might be impossible. This could lead to a skew in our available data where larger ampltiudes are more likely to be reported.

To explore the limits of LSST’s rotational sensitivity, we consider two extreme rotational populations: super-fast rotators (SFRs) and super-slow rotators (SSRs). SFRs are small Solar System bodies that rotate faster than the critical period a rubble-pile asteroid could survive without internal cohesion. Measuring these rapid rotation rates provides insight into internal strength, cohesion, and composition. In contrast, SSRs rotate on such long timescales that their light-curve variations unfold over many nights. These objects may record long-term rotational evolution driven by processes such as YORP effect \cite[e.g.,][]{vetal15} or follow from split of tidally evolved binary systems \citep[e.g.,][]{nes2020}. Among the 111 Hildas observed by the Kepler Space Telescope's K2 mission, about 18\% exhibited rotational periods greater than 100 hours, indicating super-slow rotators may be more common among Hildas than in other asteroid populations \citep{k2_ssr}. In comparison, the Transiting Exoplanet Survey Satellite (TESS) in its DR1 has observations of 26 Hildas with 17 reliably determined rotation periods, among which there are only 2 greater than 100 hours ($\sim 12\%$) \citep{2025A&A...693A..66V}. Although lower, this fraction is broadly consistent with the K2 result given the smaller sample size. Rubin's wide-field and ten-year cadence is well suited for characterizing these slow rotators, offering more complete coverage than narrow state-of-the-art surveys such as the DECam Ecliptic Exploration Project \citep{deep_sfr}. 

While real small-body light curves often contain higher-order structure from irregular shapes, surface variations, or binarity, this complexity is difficult to model in detail. We therefore adopt a simple sinusoidal model that does not include these higher-order features. Although the observed Hilda population shows a high binary fraction \citep{k2_ssr}, this approximation provides a simple framework in which we can evaluate recovery efficiency purely as a function of light-curve amplitude and rotational period. Using this simplified model which does not take into account pole orientation, we simulate Rubin's detection of Hildas to evaluate its ability to recover rotational periods. 

In Section 2 we describe our methods, including the model Hilda population, the \texttt{Sorcha} simulation of LSST. Section 3 presents our results including expected discovery yields and light-curve recovery. Section 4 provides the conclusions, literature discussion, and future work. 

\section{Methods and Inputs}
\subsection{Orbital, Size and Color Distributions}

We construct a synthetic Hilda population of orbits, sizes and colors. We sample 485,807 independent and bias-corrected orbits and magnitudes from \citet{david&david}, which reproduces both the background population and the three major collisional families. We assigned two \textit{g–r} colors (0.51 and 0.64) for the red and less–red subpopulations. The remaining color indices ($u-r$, $i-r$, $z-r$, $y-r$) were sampled from the magnitude-dependent Trojan distributions of \citet{jake_lsst}, assuming the Hildas share the same bimodality as the Jupiter Trojans \citep{wong_17_trojan_colors}. We set a constant phase slope of $G = 0.15$ \citep[see review of the $H-G$ photometric system in][]{bowell1989,mui2010}, though in reality this parameter likely varies with composition, surface texture and wavelength. A more complete treatment of phase-angle effects could improve the accuracy of the simulated photometry, especially for objects observed over a large range of phase angles. 

\subsection{Rotational Inputs}

Building on previous \texttt{Sorcha} simulation work, we include rotational behavior for our simulated objects. We assign each synthetic Hilda a sinusoidal light curve to simulate rotational variability, allowing us to test how Rubin’s cadence and depth affect the recovery of rotation periods across a range of amplitudes. To probe how Rubin's cadence and depth affect the recovery of rotational periods as a function of both period and amplitude, we model three light-curve populations: a realistic baseline population, super-fast rotators with periods shorter than 3 hours \citep{fossil_sfr} and super-slow rotators with periods longer than 100 hours \citep{k2_ssr}

\begin{figure*}
    \begin{center}
        \centerline{\includegraphics[height=0.35\textheight,width=.75\linewidth]{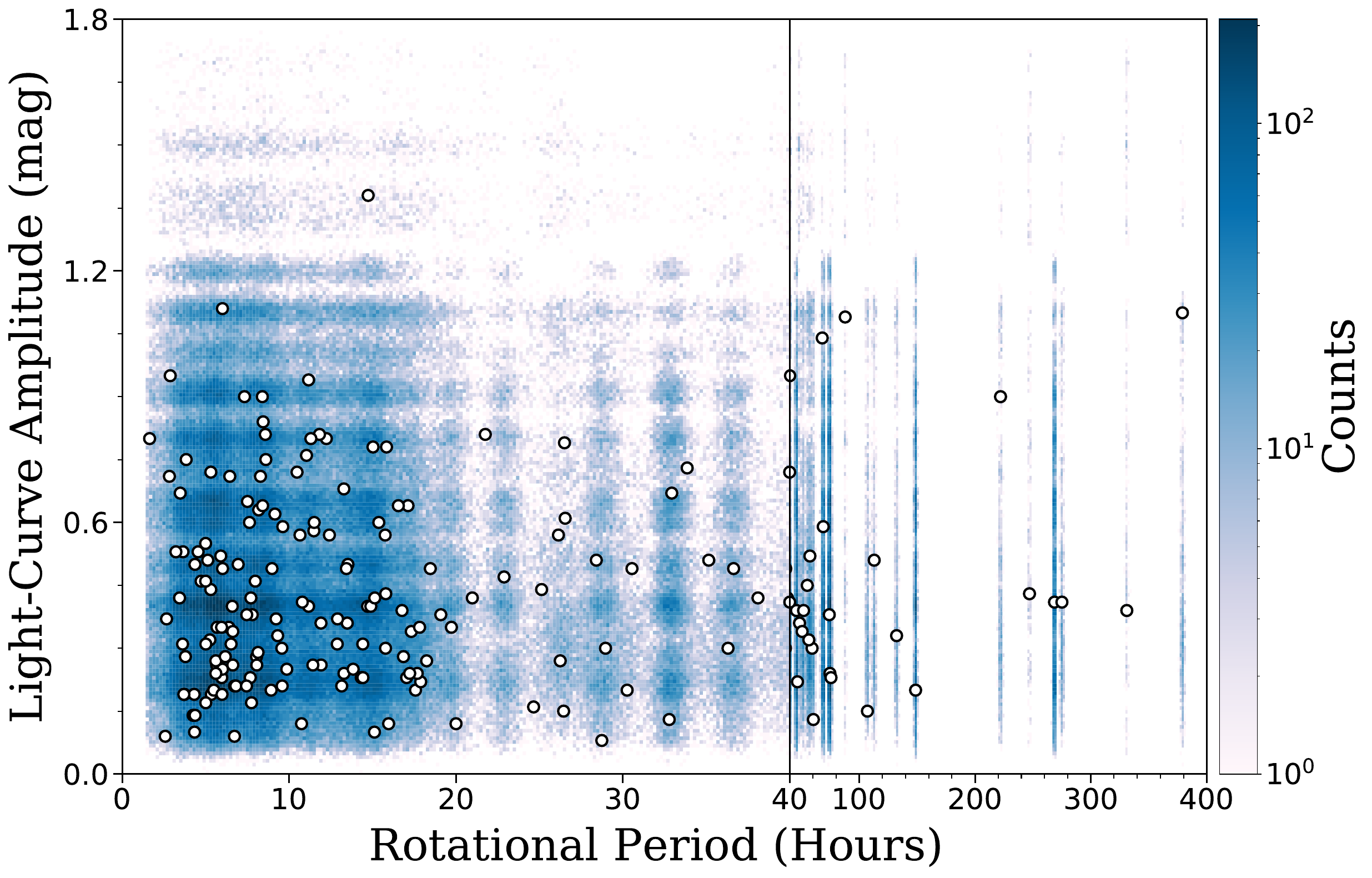}}
    
        \caption{Comparison of simulated and observed Hilda light-curve distributions. This shows a 2D histogram of rotational period versus light-curve amplitude for the combined simulated population, created by sampling periods and amplitudes from the KDE associated with each object's collisional family. White points indicate Hildas from the LCDB catalog. Our KDE-based sample is consistent with the rotational properties of observed Hildas.}

        \label{fig:realheatmap}
    \end{center}
\end{figure*}

Our baseline population is based on the 196 Hildas in the LCDB with measured light-curve amplitudes and rotational periods. Although subject to observational bias, it remains the most comprehensive source of Hilda light-curve data. The distributions of rotational periods and light-curve amplitudes were modeled using a Gaussian kernel density estimate (KDE) of the LCDB Hilda observational data. Each KDE generates a smooth, continuous function, from which we can sample arbitrarily many periods and amplitudes. Separate KDEs were generated for periods and amplitudes for each collisional family: Hilda, Schubart and the background population. 120 of these LCDB Hildas are within the Hilda resonance but are not confirmed members of any collisional family. The Potomac family had no Hildas with measured light curves in the LCDB, so we assign their rotational properties from the Hilda collisional family which shares similar orbital parameters. We adjusted the bandwidth for each KDE to obtain the widest distribution that did not spread the bulk of periods above 20 hours, below which roughly 70\% of the observed Hilda periods lie. The bandwidth values used for each collisional family are listed in Table~\ref{tab:input_pop}. This choice results in some gaps and discretization at longer periods. Using these KDEs, we generate a combined simulated Hilda population by assigning each object a period and amplitude drawn from the KDE of its collisional family. This synthetic population reflects the mixture of families in our baseline sample and reproduces the trends seen in the observed LCDB distribution (Figure~\ref{fig:realheatmap}). We identically and independently draw each object's amplitudes and periods from its collisional families period and amplitude KDEs. We believe the KDE-derived distributions are representative of the observed Hilda population in the LCDB, though these distributions are not necessarily representative of the intrinsic population.

To complement our KDE-based LCDB population, we also construct extreme input period and amplitude distributions to focus on the most exciting objects while probing potential systematic biases and blind spots in Rubin’s cadence. These distributions are not intended to represent the physical Hilda population, but rather to test the full range of periods and amplitudes that may be underrepresented or absent in the LCDB. We defined our SFR boundary following \citet{fossil_sfr}, from the critical 3-hour limit for objects with bulk density of $\sim 1.5$ g cm$^{-3}$, though main-belt observations include extreme cases with periods as short as 0.21 hours \citep{deep_sfr} and now faster by \citet{Greenstreet_RFL}. We set the SSR lower bound to 100 hours following \citet{k2_ssr} and its upper bound to 1400 hours—roughly twice the largest period among the 1,465 LCDB Hildas. For each population, we drew rotational periods uniformly: from 0–3 hours for SFRs and from 100–1400 hours for SSRs. We sampled amplitudes uniformly between 0 and 2.5 mag, approximately twice the largest observed Hilda amplitude in the LCDB, to account for possible highly-elongated shapes. Together, these extreme populations enable us to explore detectability limits and identify potential systematic biases arising from Rubin’s cadence, complementing our more realistic LCDB-based rotational population.

\subsection{Sorcha Simulations}

We simulate LSST observations of Hildas using \texttt{Sorcha}, a solar system survey simulator designed for large-scale wide-field surveys like LSST. Given a survey’s cadence, field coverage and a model Solar System population, \texttt{Sorcha} accurately integrates orbits \citep{Sorcha_matt} and rotational light curves for each object and evaluates their detectability in each exposure. Detections brighter than the $\sim 16.0$ mag saturation limit are excluded, as they are not reliably measured. 

We adopt the latest Rubin-published (v4.2) baseline LSST cadence \citep[][as of July 17, 2025;]{yoachim23,opsim_2014, peter_yoachim_2025_14847371}, which provides a realistic model of the telescope's observing pattern. This ten-year simulated survey begins January 1, 2026 and emphasizes the wide-fast-deep strategy, supplemented by deep-drilling fields and the Northern Ecliptic Spur to maximize Solar System discovery. Observing conditions such as filter, sky brightness and limiting magnitude are also included so that the simulated detections accurately reflect LSST's observing capabilities.  For each detection of each object in each image, \texttt{Sorcha} provides the photometric measurements we use to construct light curves for our recovery tests. We adopted \texttt{Sorcha}’s default parameters and ran all simulations on the UW Epyc cluster, requiring roughly 100 core hours in total.

\subsection{Period Recovery}

To test whether we could recover rotational periods from simulated LSST observations, we applied a multiband Lomb-Scargle periodogram via \texttt{Astropy} \citep{astropy:2022} to the simulated light curves. The Lomb–Scargle periodogram produces a power spectrum across a range of frequencies. The multiband implementation adds a color offset parameter for each optical filter \citep{deep_sfr, multiband_ls}, allowing data from each filter to contribute to the determination of a single period.

We limited the frequency search window to the bounds of each input population. For our SFRs, we limited our search window to $0.2 - 3$ hours. Periods approaching zero hours require increasingly dense frequency sampling and substantially increase computational runtime, and a lower limit of 0.2 hours allows for the recovery of the 0.21-hour period main belt asteroid \citep{deep_sfr}. For SSRs our search window spans 24 to 1400 hours to match the input population and to allow objects to be fit to the common 24-hour alias. These frequency windows are intentionally restricted and do not represent the full physical range of rotational periods, but make period recovery computationally feasible. We set the Lomb-Scargle samples-per-peak parameter, which controls the frequency resolution, to 50 for each period search \citep{VanderPlas_2018}.

\begin{figure}
    \begin{center}
    \hspace{10mm}
     \centerline{\includegraphics[height=0.28\textheight,width=1.1\linewidth]{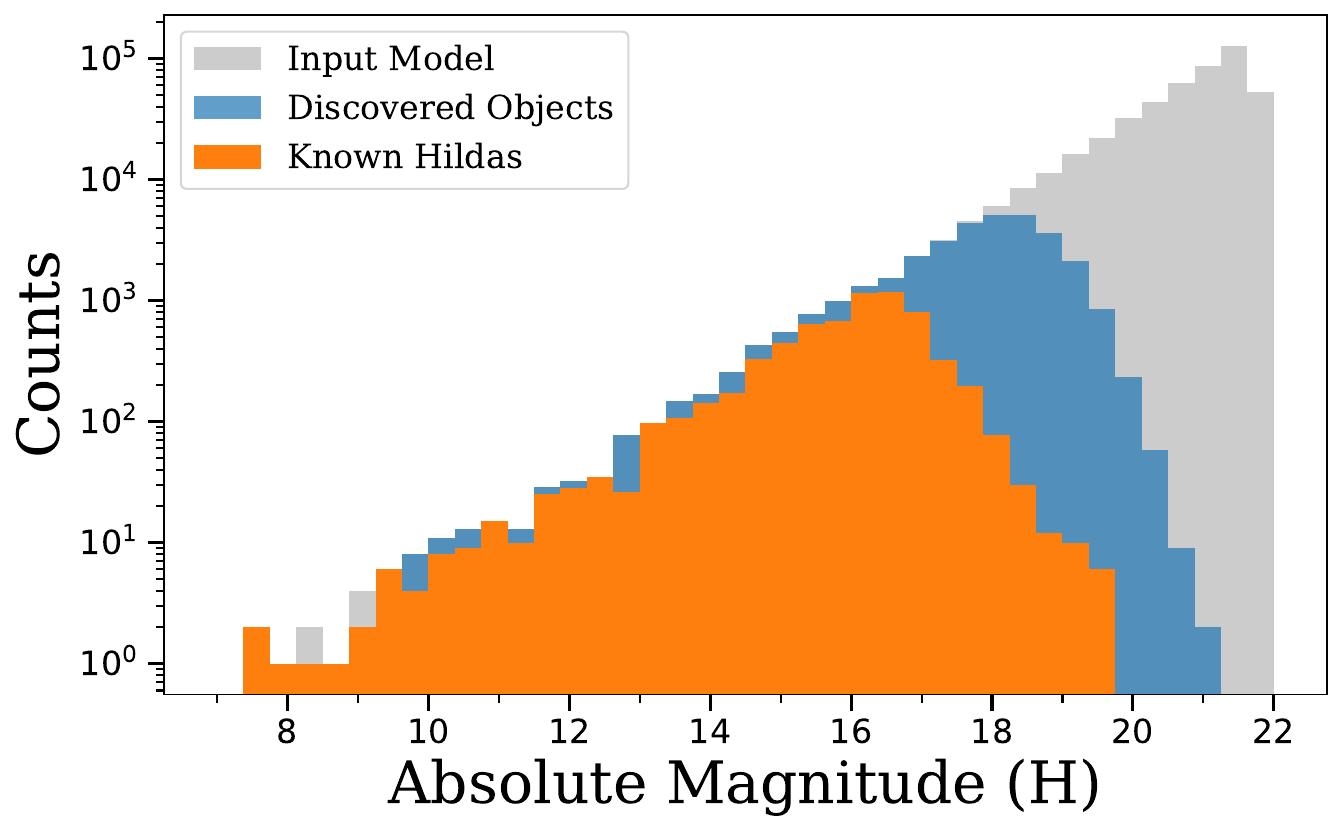}}
    
        \caption{Comparison of absolute magnitude ($H$) distributions for the known, input and discovered Hilda populations. The orange histogram shows the currently-known Hildas from the MPC. The blue distribution represents the Hildas discovered in our simulated survey with the gray being the full input population. The synthetic discoveries extend $\sim$1.5–2 magnitudes fainter than the known population, demonstrating LSST’s ability to probe substantially deeper and expand the observed Hilda population beyond current detection limits.}

        \label{fig:mpcdiscovered}
    \end{center}
\end{figure}

\begin{figure*}
    \begin{center}
    \hspace{100mm}
     \centerline{\includegraphics[height=0.31\textheight,width=1\linewidth]{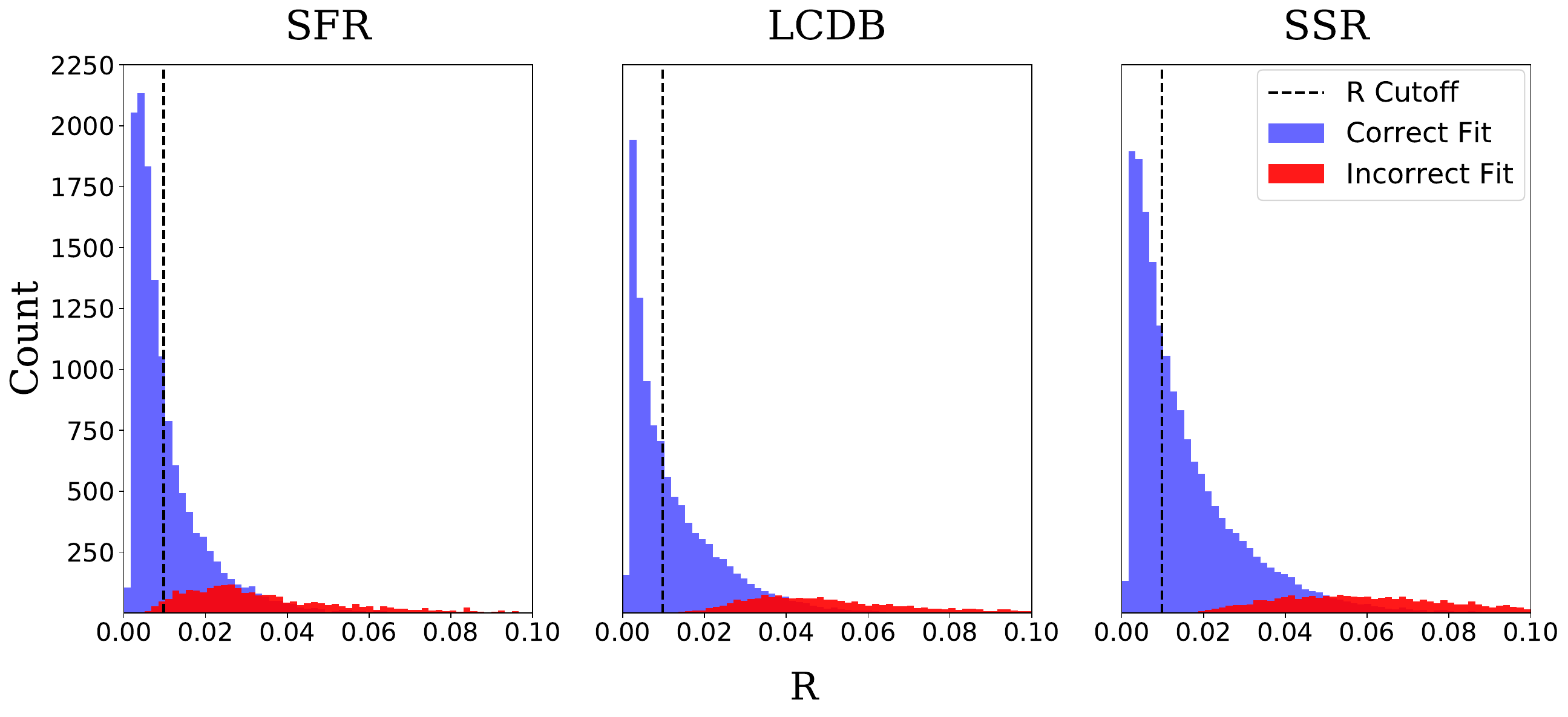}}
    
        \caption{Distributions of the reduced inverse power metric $R$ (Equation \ref{reduced_inv_pow}) for the SFR, LCDB and SSR simulated light–curve populations, separated into the correctly and incorrectly recovered rotational periods. The vertical dashed line shows the confidence threshold $R = 0.01017$, defined as the value below which at least 99\% of SFR periods are accurately recovered (including half/double-period harmonics). Correct fits (blue) concentrate strongly at low $R$, while incorrect fits (red) populate a broader tail toward larger $R$, demonstrating that smaller $R$ values correspond with more reliable period recoveries.}
        
        \label{fig:R_cutoff}
    \end{center}
\end{figure*}

For each simulated light curve, the periodogram identified the period associated with the highest-power peak. To assess the reliability of each period fit, we introduce a ``reduced inverse power" confidence metric. For each object we identify the highest periodogram power $P_{\rm max}$ across all frequencies and compute:
\begin{equation}
    R = \frac{1}{P_{\rm max}\,(N_{\rm obs}-N_{\rm Parameters})}~,
    \label{reduced_inv_pow} 
\end{equation}
where $P_{\rm max}$ is the maximum periodogram power and $N_{\rm obs}$ is the number of observations and $N_{\rm Parameters}$ represents the number of parameters in our model. Here, $N_{\rm Parameters}~ = ~9$, accounting for the sinusoidal terms, period and the multiband offset parameters. We then compare our best period fit with the input period for each object and define an accurate fit as being within 1\% of the true input period (including half and double harmonics). We define our confidence threshold at the value of $R$ below which 99\% of periods are accurately recovered; this yields $R= 0.01017$. This represents the strictest cutoff among the three simulated populations and we apply it for all populations for consistency. The distribution of $R$ values for correct and incorrect periods is shown in Figure~\ref{fig:R_cutoff}. Lower $R$ values correspond to higher confidence, with most accurate period recoveries concentrated at low $R$ and incorrect periods increasingly likely at higher $R$.

\section{Simulated LSST Yield For Hildas}
\subsection{Discovery Yield and Completeness}
\begin{figure*}
    \begin{center}
    \hspace{100mm}
     \centerline{\includegraphics[height=0.32\textheight,width=1\linewidth]{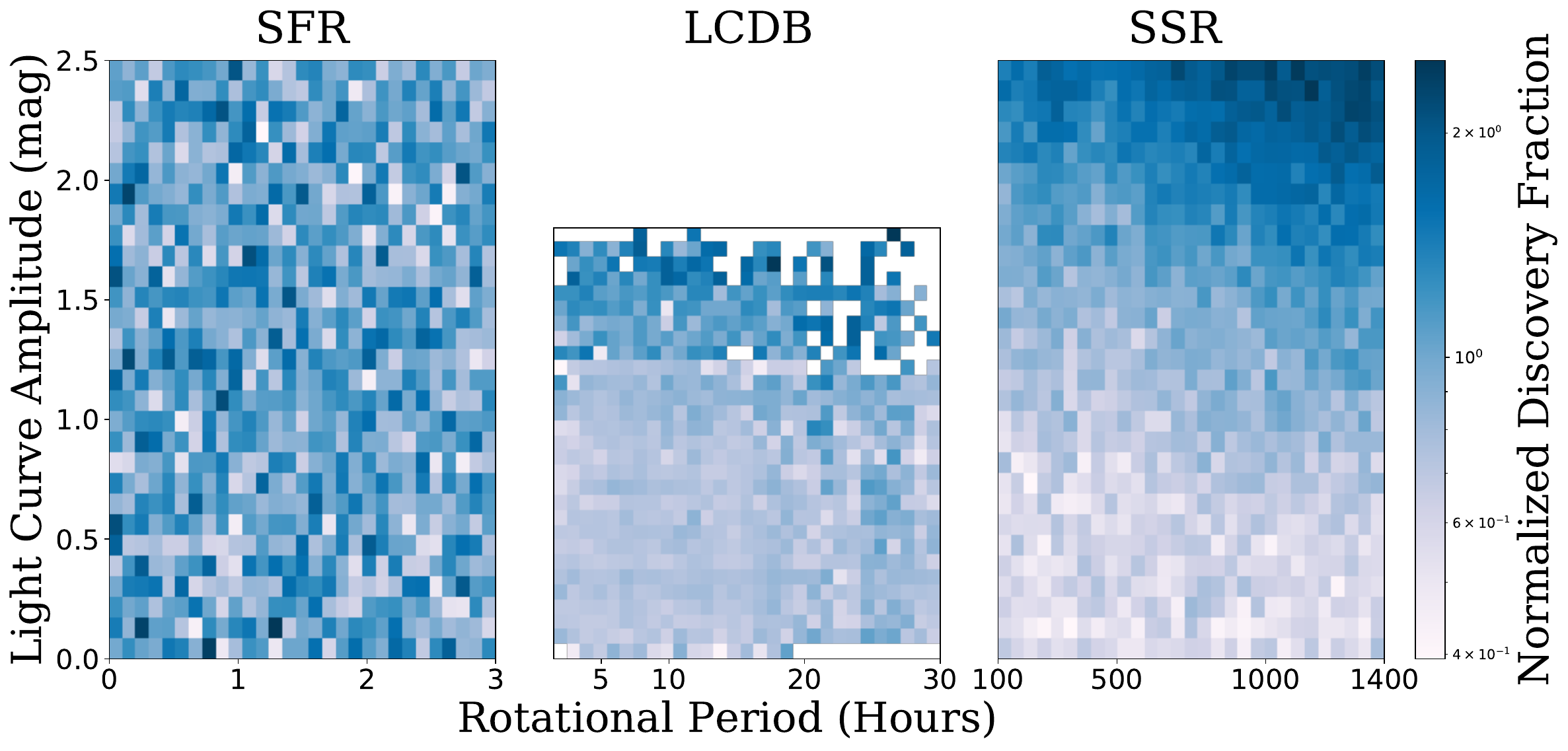}}
    
        \caption{Discovery fraction of our simulated Hilda bodies for the SFR, LCDB and SSR populations. Each panel displays the per-bin fraction of objects that were detected in our simulated survey, normalized by the overall discovery fraction so that an average bin has a value of 1.0.}
        
        \label{fig:discover_frac}
    \end{center}
\end{figure*}

Our simulated LSST survey discovers 33,405 Hildas in the baseline population. This represents a factor-of-five increase over the 6,622 currently known Hildas \citep{JPL_SBDB} (Figure~\ref{fig:mpcdiscovered}). The simulated survey has very high discovery completeness for bright Hildas. Completeness is 100\% for Hildas with $H_r$ between 9.2 and 17.0 and remains above 90\% for $H_r$ up to 18.0. The faintest object detected has an $H_r$ of 21.0 with the faintest Hilda in the MPC database of 19.7 magnitudes. Completeness drops at the bright end due to the $m_r \approx $ 16.0 due to saturation and at the faint end due to the surveys limiting depth. The average number of observations of discovered Hildas in the baseline population is 173.6. 73 objects were detected at least 1000 times each, with the most-detected object having 3769 detections. For comparison, K2 observations of Hildas ranged from 218 to 1596 detections per object with 10 of 102 Hildas listed having greater than 1000 detections \citep{k2_ssr}. LSST is expected to reveal tens of thousands of faint members of the Hilda population that were previously undetectable by past surveys.

\subsection{Light-Curve Recovery}

\begin{figure*}
    \begin{center}
    \hspace{100mm}
     \centerline{\includegraphics[height=0.32\textheight,width=1\linewidth]{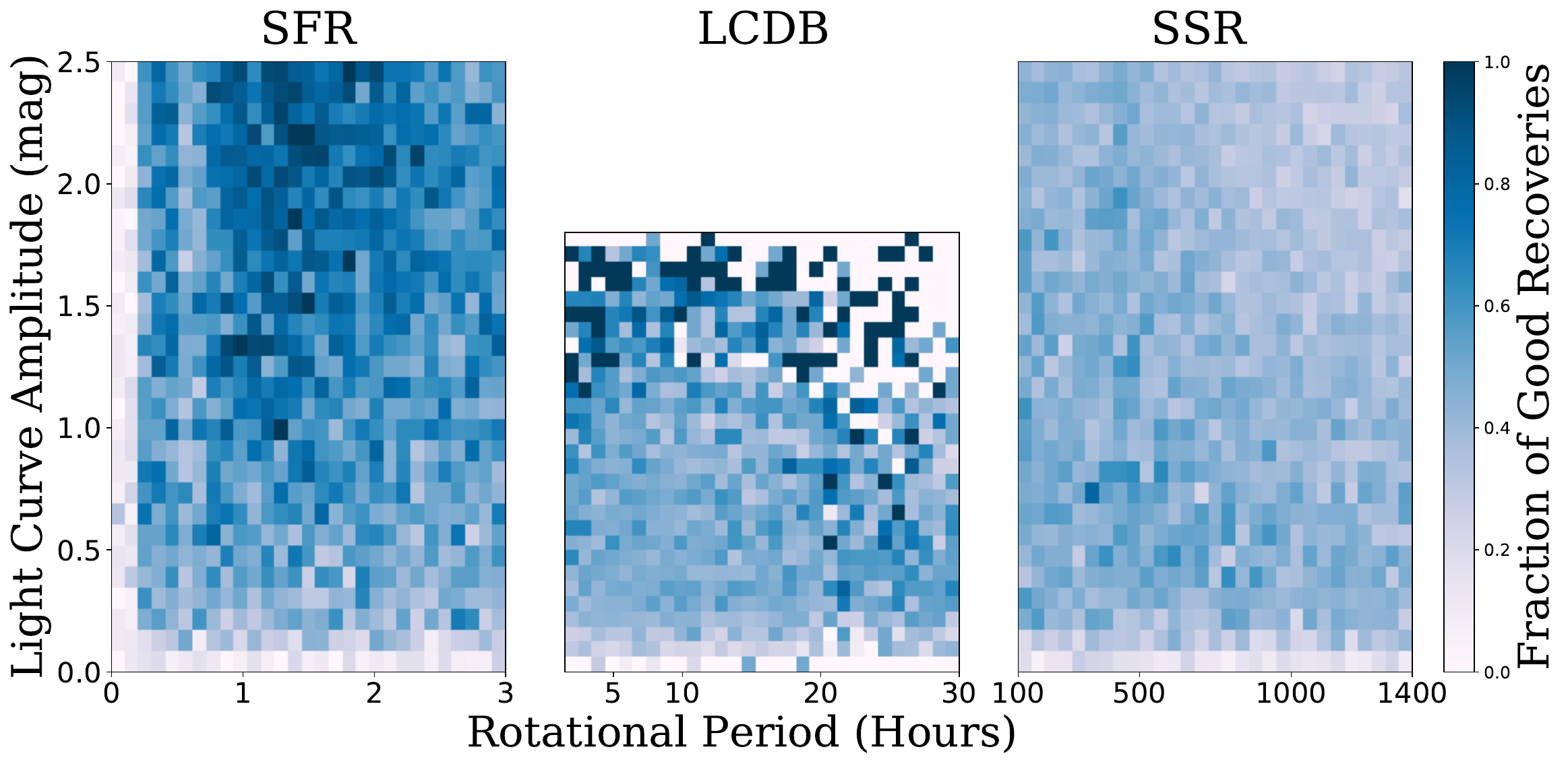}}
    
        \caption{Fraction of light curves meeting the 99\% confidence threshold as a function of light-curve amplitude and rotational period. Each bin shows the fraction of objects whose Lomb–Scargle periodogram structure indicates high confidence in the recovered period. Recovery confidence decreases significantly at lower amplitudes, where weaker variability reduces the reliability of period identification.}

        \label{fig:periodrecovery}
    \end{center}
\end{figure*}

We are highly confident in our recovered periods for 15,500 of 33,405 (46.5\%) Hildas in our LCDB simulation population (Figure~\ref{fig:periodrecovery}). Period recovery efficiency is largely determined by light-curve amplitude. For light-curve amplitudes greater than 0.2 mag, recovery generally exceeds 45\%. Among the uniform input populations, we achieved roughly 9\% high-confidence recovery for amplitudes from 0 to 0.05 magnitudes. For the SSR population, we see a peak of 45\% confidence around 0.5 magnitudes before a steady decline in recovery rate to roughly 35\% at 2.5 magnitudes. This decline largely reflects the limited number of detections for very faint, long-period objects. Many of these bodies remain detectable for only a portion of the survey window, so their light curves are sparsely sampled, reducing the likelihood of successful period recovery. We also note that only two periods (searched from 24 to 1400 hours) were folded to the 24 hour alias, and none were flagged as highly confident recoveries.

For the SFR population, period recovery is generally higher compared to the other two populations, most likely due to its larger fraction of high-amplitude light curves compared to the LCDB population. Recovery rates are typically around 65\% but have higher variance than the other populations. We see a characteristic dip in recovery efficiency near LSST's typical revisit cadence of $\sim$ 36 minutes and its harmonic at 18 minutes. On either side of this cadence-related dip, the recovery rate exceeds 60\%, while within the dip the rate falls to $\sim$ 45\% (Figure~\ref{fig:periodrecovery}). 

For the SSR population, recovery fraction decreases with rotational period. For periods from 100 hours to 200 hours, the recovery confidence is approximately 45\%, and remains around 40\% until approximately 700 hours. Beyond 700 hours, recovery confidence steadily declines to roughly 35\% at extremely long periods, near 1300 hours. In Figure~\ref{fig:periodrecovery}, compared to the LCDB and SFR populations, the SSR population shows slightly higher recovery fractions at intermediate amplitudes (0.3–1.0 mag) and shorter periods ($<$ 700 hours), but lower discovery rate, likely because objects with high amplitudes (2.0-2.5) and long periods ($>$ 700 hours) spend more time near consecutively near maximum brightness, increasing the chance of being detected in three tracklets in a row (Figure~\ref{fig:discover_frac}). However, many of these objects are very dim and sparsely observed, so their light curves remain poorly sampled, which reduces the likelihood of accurate period recovery compared to brighter, constantly detected objects. Only 23 periods (0.04\%) in the SSR population are affected by the 24-hour nightly alias. Among these, nearly 70\% correspond to input periods exceeding 1000 hours and the average number of detections was below 20, suggesting that extremely slow rotators which are observed less frequently are particularly susceptible to being folded to this daily timescale. None of these folded objects meet our confidence threshold, so we safely label all of these aliased periods as unreliable. 

Variations in surface properties and light-curve amplitudes among Hildas can strongly influence our ability to measure rotational periods. We assume a constant phase slope ($G = 0.15$) and uniform surface properties only crudely represent the diverse surfaces of these objects. Most importantly, a large majority of our simulated objects have light-curve amplitude that can be measured; 97.9\% of our input Hildas having amplitudes greater than 0.1 magnitudes, while the intrinsic population could possibly have amplitudes far lower than this. The Minor Planet Center lists 6,622 known Hildas currently and the LCDB lists light-curve parameters for 196 of them, suggesting that many Hildas may have low amplitudes, making accurate period recovery significantly more difficult. 

\section{Discussion}
\subsection{Conclusion}

Using the \texttt{Sorcha} survey simulator, we produced a high-fidelity simulation of LSST observations for the Hilda asteroid population. Our synthetic survey predicts a roughly fivefold increase in the known population and shows that LSST will discover all bright Hildas with $H_r$ between 9.2 and 17.0.

We are highly confident in the recovery of 46.5\% of Hildas in our LCDB-based population. Period recovery, assessed using a multiband Lomb-Scargle periodogram, is shaped strongly by objects light-curve amplitude. High-amplitude objects are more likely to be confidently recovered. For light-curve amplitudes greater than 0.2 mag, recovery generally exceeds 45\%. Among the uniform input populations, we achieved roughly 9\% high-confidence recovery for amplitudes from 0 to 0.05 magnitudes. We assume a constant phase slope ($G = 0.15$) and uniform surface properties which might not accurately represent the complex surfaces of these objects. Most importantly, a large majority of our simulated objects have light-curve amplitude that can be measured: 97.9\% of our input Hildas have amplitudes greater than 0.1 magnitudes, while the intrinsic population might have lower typical amplitudes.

\subsection{Literature Discussion}

A key limitation of our light-curve model is the absence of rotation-pole orientations. In our simulation, each object has a fixed sinusoidal light curve with a constant amplitude, equivalent to an orientation which maximizes the observable magnitude variation. The observed Hilda population shows some concentration of pole obliquities near $0^{\circ}$ and $180^{\circ}$ \citep{dh2023}, but a substantial number occupy intermediate obliquities, which affect the recoverability of their light curves. Similarly, real objects have diverse phase slopes, introducing additional brightness variations not captured in our constant-phase model. While the effect of pole obliquity on light-curve recovery is uncharacterized, these assumptions mean the current period recovery rate is likely an upper limit.

The large number of repeated measurements provided by LSST will enable the construction of a bias-corrected catalog of true rotational properties for the Hilda population. For the majority of objects, hundreds of measurements will be obtained, with the most-observed objects receiving more than 1,000 detections. We can use survey simulations to account for observational biases as a function of rotational period and amplitude. Comparisons between the debiased rotational states of different collisional families will provide insight into collisional physics, material-dependent non-gravitational effects such as YORP and the formation and dynamical evolution of the Hilda population. 

Our results provide an initial benchmark for LSST’s capabilities of Hildas and are expected to scale to other populations with similar observational properties. The discovery and period-recovery fractions for Hildas should be broadly comparable to those of Jupiter Trojans and main-belt asteroids, given their similar brightness distributions and LSST observing cadence. Our simulated catalogs and methodology provide a framework for interpreting early LSST light-curve data, enabling future studies to compare period-recovery and selection effects against a well-characterized synthetic population. This work extends previous Solar System survey simulations \citep{jake_lsst} to the Hilda population and light-curve recovery. 

\citet{Greenstreet_RFL} analyzed Rubin Observatory First Look data and performed light-curve period and amplitude fitting using a method similar to ours. Their search was limited to a narrow 14-square degree field over 12 days, but the dozens of consecutive repeat measurements in their cadence make light-curve recovery far more efficient than the LSST ten-year baseline simulation. Our simulation with the ten-year LSST cadence produces an average of over 170 detections per object, similar to the roughly 150 detections per object in the Rubin First Look sample. However, unlike the real RFL data, our simulation relies on simplified assumptions about light-curve amplitude, phase slope, and uniform surface properties, which may lead to overestimated period recoverability compared to the true Hilda population. Because RFL represents the first real LSST-like dataset, it provides an ideal testbed for applying the selection-function framework developed in this work. A natural next step is to debias the RFL rotation-state sample using our LSST-calibrated period-recovery model to infer the intrinsic distribution of asteroid rotation periods, amplitudes, and shapes.

\subsection{Future Work}

LSST data will enable a debiased characterization of rotational properties of Solar System populations. The next step is to extend this framework to real LSST observations to validate our simulated recovery rates, debias a real search for rotation periods, and construct population-specific models of the intrinsic asteroid rotation state distributions.

Several additional parameters should be incorporated into future work to refine the selection function and better simulate the intrinsic population. For example, variable pole obliquity could be modeled in \texttt{Sorcha} simulations to more realistically capture the dependence of period-recovery efficiency on spin distribution. In future work, we plan to model arbitrary pole directions and extend the \citet{Greenstreet_RFL} selection function to amplitude, period, and obliquity simultaneously, enabling LSST observations to constrain the intrinsic pole–obliquity distribution and inform models of collisional evolution and YORP-driven spin-state modification \citep{dh2023}. The multiband photometry provided by LSST also opens opportunities to explore color-dependent effects on light-curve recovery. In this study, we assigned two \textit{g-r} colors (0.51 and 0.64) while keeping remaining color bands (\textit{i-r, z-r, y-r}) constant. We did not test whether light-curve recoverability differs between these two classes. Future work could take into account the population-dependent period-recovery efficiencies of the color classes and use them to investigate how spin-state properties vary with bulk density, including potential differences in the spin barrier. LSST may also enable testing whether collisional families exhibit distinct rotational properties, providing new constraints on Hilda formation, evolution, and broader early–Solar System processes.

\section{Acknowledgments}
This work made use of the following software packages: \texttt{astropy} \citep{astropy:2013,astropy:2018,astropy:2022}, \texttt{Jupyter} \citep{2007CSE.....9c..21P,kluyver2016jupyter}, \texttt{matplotlib} \citep{Hunter:2007}, \texttt{numpy} \citep{numpy}, \texttt{pandas} \citep{mckinney-proc-scipy-2010,pandas_17229934}, \texttt{python} \citep{python}, and \texttt{scipy} \citep{2020SciPy-NMeth,scipy_17467817}. Software citation information aggregated using \texttt{\href{https://www.tomwagg.com/software-citation-station/}{The Software Citation Station}} \citep{software-citation-station-paper,software-citation-station-zenodo}.

A.J.F. thanks J.A.K for his mentorship and guidance throughout this project.
A.J.F. and J.A.K. acknowledge support from the University of Washington College of Arts and Sciences Department of Astronomy. J.A.K thanks the LSST-DA Data Science Fellowship Program, which is funded by LSST-DA, the Brinson Foundation, and the Moore Foundation; his participation in the program has benefited this work.
J.A.K., P.H.B., and M.J. acknowledge support from the DIRAC Institute in the Department of Astronomy at the University of Washington. The DIRAC Institute is supported through generous gifts from the Charles and Lisa Simonyi Fund for Arts and Sciences, and the Washington Research Foundation. 
This material is based upon work supported by the National Science Foundation under Grant No. (2307569). 
This research award is partially funded by a generous gift of Charles Simonyi to the NSF Division of Astronomical Sciences. The award is made in recognition of significant contributions to Rubin Observatory’s Legacy Survey of Space and Time. Any opinions, findings and conclusions or recommendations expressed in this material are those of the author(s) and do not necessarily reflect the views of the National Science Foundation.
The work of DV was partially supported by the Czech Science Foundation (grant 25-16507S).

\begin{table*}[ht!]
\centering
\renewcommand{\arraystretch}{1.2}
\caption{Input Parameters of Simulated Hilda Populations}
\begin{tabular}{@{\hskip 0.3cm} l @{\hskip 2.5cm} r @{\hskip 0.3cm}}
\hline
\textbf{Parameter} & \textbf{Input Value / Parameter} \\
\hline
Range of semi-major axis $(a)$ & 3.8858–4.0396 (au)\\
Range of eccentricity $(e)$ & 0.0012–0.3905 \\
Range of inclination $(i)$ & 0.015–21.69 (deg)\\
Phase slope parameter ($G$) \rule{0pt}{2.5ex} & 0.15 \\
Range of Absolute magnitude ($H_r$) & 7.53–21.75 (mean = 20.63) \\
Color ($g - r$) & 0.51 (483,215 objects), 0.64 (2,592 objects) \\
Color index ($u - r$) & 1.4353 \\
Color index ($i - r$) & $-0.22$ \\
Color index ($z - r$) & $-0.39$ \\
Color index ($y - r$) & $-1.1804$ \\
Phase function model & $HG$ \\
Light-curve model & Sinusoidal \\
Period KDE bandwidth (Hilda / Schubart / Background) & 0.0045 / 0.008 / 0.01 \\
Amplitude KDE bandwidth (Hilda / Schubart / Background) & 0.07 / 0.05 / 0.08 \\
\hline
\end{tabular}
\vspace{0.15cm}
\\
\textit{Note---}Orbital elements and absolute magnitudes are derived from \citet{david&david} and colors from \citet{wong_17_trojan_colors}.
\label{tab:input_pop}
\end{table*}

\bibliography{my_refs.bib}

\end{document}